\documentclass[12pt]{iopart}
\usepackage[english]{babel}
\usepackage[utf8]{inputenc}

\usepackage{iopams}
\usepackage{float}
\usepackage{graphicx}
\usepackage{multirow}
\usepackage{pdflscape}
\usepackage{multirow}

\usepackage{bm}
\usepackage{hyperref}
\hypersetup{
colorlinks,
citecolor=blue,
filecolor=blue,
linkcolor=blue,
urlcolor=blue}

\usepackage{subfigure}

\newcommand{\la}{\left\langle}
\newcommand{\ra}{\right\rangle}
\newcommand{\be}{\begin{equation}}
\newcommand{\ee}{\end{equation}}
\newcommand{\bea}{\begin{eqnarray}}
\newcommand{\eea}{\end{eqnarray}}
\newcommand{\ba}{\begin{array}}
\newcommand{\ea}{\end{array}}

\usepackage{makeidx}
\makeindex 
\begin{document}
\title{Contrasting turbulence in stably  stratified flows and  thermal convection}

\author{Mahendra K. Verma}

\address{Department of Physics, Indian Institute of Technology Kanpur, Kanpur 208016, India}
\vspace{10pt}
\begin{indented}
\item[]August 2016
\end{indented}

\begin{abstract}

In this paper, the  properties of  stably stratified turbulence (SST) and turbulent thermal convection are contrasted.  A key difference between these flows is the sign of the kinetic energy feed by buoyancy, $\mathcal{F}_B$.   For SST, $\mathcal{F}_B < 0$ due to its stable nature; consequently, the kinetic energy flux $\Pi_u(k)$ decreases with wavenumber $k$ that leads to a steep kinetic energy spectrum, $E_u(k) \sim k^{-11/5}$.  Turbulent convection is unstable, hence  $\mathcal{F}_B > 0$  that leads to an increase of $\Pi_u(k)$ with $k$; this increase however is marginal due to relatively weak buoyancy, hence $E_u(k) \sim k^{-5/3}$, similar to that in hydrodynamic turbulence.  This paper also describes the conserved fluxes for the above systems.

\end{abstract}

\section{Introduction}

Buoyancy affects the flows in planets and stars, hence its understanding is very critical.   Buoyancy-driven flows come in two categories: (a) Stably stratified flows, as in Earth's atmosphere; (b) Unstably stratified flows, as in thermal convection, Raleigh-Taylor instability, etc.  These two classes of flows have very different behavior, e.g., nature of  isotropy and large-scale flows, spectral energy spectrum and flux, etc.    The above topics have extensive literature~\cite{Ahlers:RMP2009,Chilla:EPJE2012,Davidson:book:TurbulenceRotating,Lohse:ARFM2010,Verma:book:BDF}, and even an introduction of these topics would take many pages.   Hence this paper is limited  to contrasting the energy spectrum and flux of the aforementioned two classes of flows.   In addition, among the unstably stratified flows, I focus on turbulent thermal convection.

\begin{figure}[htbp]
\begin{center}
\includegraphics[scale = 0.6]{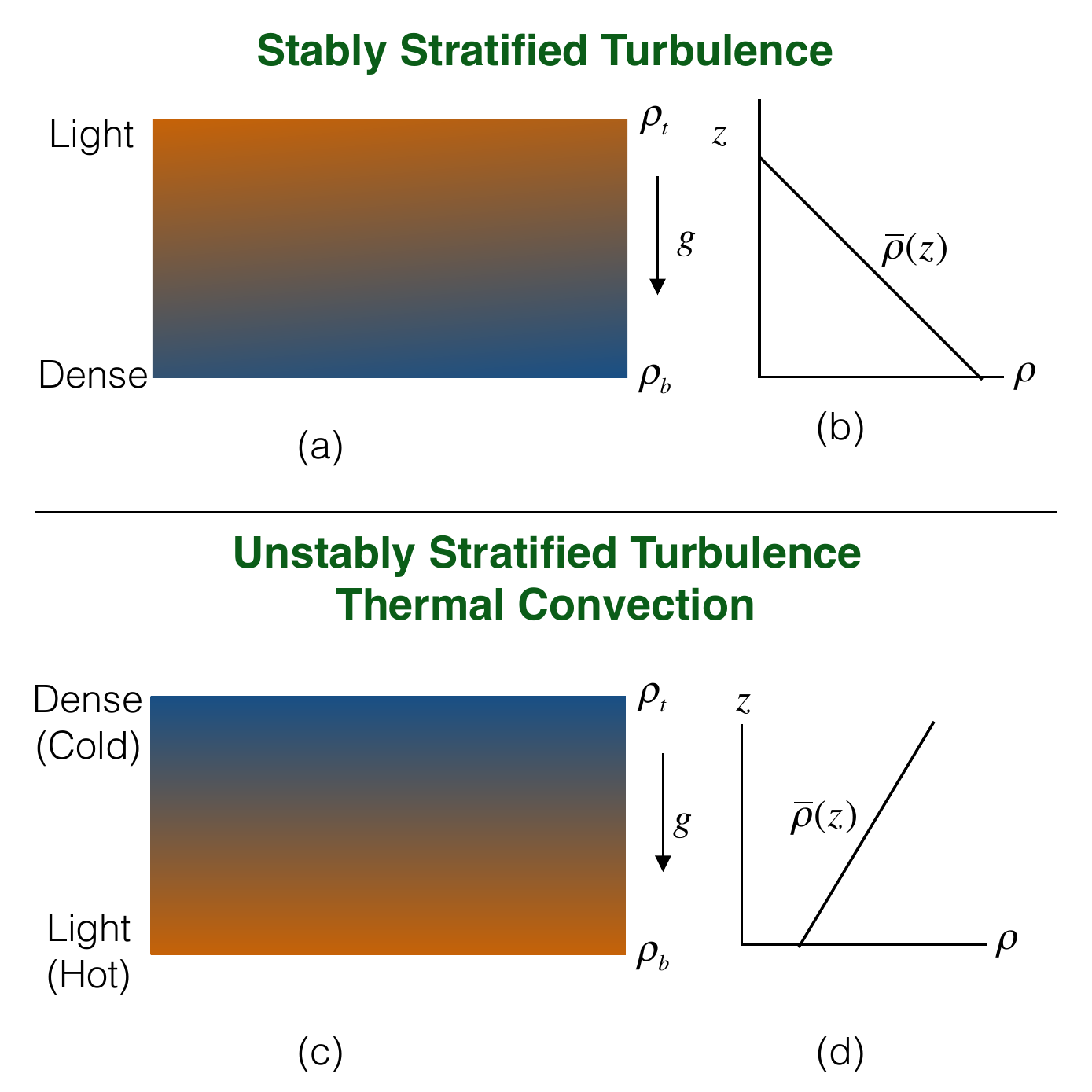}
\end{center}
\caption{(a) Schematic diagram of a stably stratified flow:  Lighter fluid is above the heavier fluid. $\rho_b$ and $\rho_t$ are the fluid densities at the bottom and top layers  respectively.  (b) The mean density $\bar{\rho}(z)$ decreases linearly with $z$. (c, d) Schematic diagram of  thermal convection in which the density variation is reversed. The bottom plate is hotter than the top plate. }
\label{fig:density_schematic}
\end{figure}

Figure~\ref{fig:density_schematic}  exhibits schematic diagrams of a stably stratified flow and thermal convection.  In stably stratified flows, the density decreases with vertical height.  The density variation however is reversed in thermal convection.  These density stratifications make the flow stable and unstable respectively.   These contrasting nature of stability leads to very different energy spectra for the two sets of flows.

The energy spectra of stably stratified flows depend critically on the strength of stratification.  This paper is focussed on moderate stratification for which the flow is nearly isotropic.   For such flows, Bolgiano~\cite{Bolgiano:JGR1959} and Obukhov~\cite{Obukhov:DANS1959} provided the first phenomenological model, referred to as Bologiano-Obuknov (BO)  phenomenology or scaling.  In this model, the kinetic energy spectrum, $E_u(k)$,  is proportional to $k^{-11/5}$, which is steeper than Kolmogorov's $k^{-5/3}$ spectrum.  The additional steepening of inertial $E_u(k)$ is because of the attrition of kinetic energy flux due to the transfer of kinetic energy to the potential energy.

Stably stratified flows and thermal convection are described by very similar equations, hence many researchers assumed that the BO phenomenology would be applicable to turbulent thermal convection as well~(\cite{Ahlers:RMP2009,Chilla:EPJE2012,Lohse:ARFM2010,Lvov:PRL1991,Lvov:PD1992,Rubinstein:NASA1994,Verma:book:BDF} and references therein).  In particular, using field-theoretic and scaling arguments,   L'vov~\cite{Lvov:PRL1991},  L'vov and Falkovich~\cite{Lvov:PD1992}, and Rubinstein~\cite{Rubinstein:NASA1994} argued that turbulent thermal convection follows  BO scaling.    A large number of experiments and numerical simulations were performed to test the above hypothesis, but they have been inconclusive with some reporting BO scaling, while others reporting Kolmogorov's $k^{-5/3}$ kinetic energy spectrum (\cite{Ahlers:RMP2009,Chilla:EPJE2012,Lohse:ARFM2010,Lvov:PRL1991,Lvov:PD1992,Rubinstein:NASA1994,Verma:book:BDF} and references therein).    In the present paper it is shown that the kinetic energy flux of turbulent convection is very different from that of stably-stratified flows due to the unstable nature of  thermal convection. Following this thread and several numerical findings, it is shown that the turbulence properties of turbulent thermal convection are very similar to those of hydrodynamic turbulence---namely Kolmogorov's $k^{-5/3}$ spectrum, and constant kinetic energy flux.

The outline of the paper is as follows: Secs.~\ref{sec:SST} and \ref{sec:turbulent convection}  introduce the governing equations, turbulence  phenomenologies, and numerical results of stably stratified flows and thermal convection respectively.     The last section contains a summary of the differences between the two flows.

\section{Stably stratified turbulence}
\label{sec:SST} 

A typical stably stratified flows  appears as in Fig.~\ref{fig:density_schematic}(a).  The local density, $\varrho(x,y,z)$, of the flow at ${\bf r} = (x,y,z)$ is
\be
\varrho(x,y,z) = \bar{\rho}(z) + \rho(x,y,z),
\label{eq:rho_l} 
\ee
where $\bar{\rho}(z)$ is the mean density at height $z$, and $\rho(x,y,z)$ is the fluctuation around this mean.  It is customary to assume a linear density profile for $\bar{\rho}(z)$:
\be
 \bar{\rho}(z)  =  \rho_b + \frac{d \bar{\rho}}{d z} z =  \rho_b + \frac{\rho_t-\rho_b}{d} z.
 \label{eq:linear_density}
 \ee

For stably stratified flows, it is convenient to express the density variable in the unit of velocity, $b  = (g \rho)/(N \rho_m)$, in terms of which the governing equations for stably stratified turbulence (SST) are
\bea
\frac{\partial{\mathbf{u}}}{\partial{t}}+ (\mathbf{u}\cdot \nabla)\mathbf{u} & = &  -\frac{1}{\rho_m} \nabla\sigma - N b \hat{z} + \nu \nabla^{2}\mathbf{u} +{\bf F}_u,  
\label{eq:u_SS} \\
\frac{\partial{b}}{\partial{t}}+(\mathbf{u}\cdot\nabla)b & = & N  u_{z} + \kappa \nabla^{2}b. \label{eq:b_SS}
\eea
where ${\bf u}, \sigma$ are the velocity and pressure fields respectively, $g$ is the acceleration due to gravity, ${\bf F}_u$ is the external force (in addition to buoyancy), $- N b \hat{z}$ is the buoyancy, and $\nu,\kappa$ are the kinetic viscosity and thermal diffusivity respectively.  The parameter
\be
 N = \sqrt{\frac{g}{\rho_m} \left| \frac{d \bar{\rho}}{d z} \right|}
 \ee
is the {\em Brunt-V\"{a}is\"{a}l\"{a} frequency}, which is related to the frequency of the internal gravity waves.

The densities of kinetic   and potential energies  are 
\be
E_u = \frac{u^2}{2};~~~E_b = \frac{b^2}{2}
\ee
respectively.    In the absence of external force ${\bf F}_u$,  $\nu$, and $\kappa$, the total energy
\be
E = \frac{1}{2} \int d{\bf r} (u^2 + b^2)
\ee
is conserved.  However,  when  ${\bf F}_u=0$ but $\nu, \kappa \ne 0$, the total energy of the dissipative SST decays.  Therefore,  an external force ${\bf F}_u$ is needed to maintain  a steady state.  In the present paper it is assumed that ${\bf F}_u$ is employed at large scales.

 Some of the relevant nondimensional parameters of SST are Reynolds number, Re, which is $u_\mathrm{rms}d/\nu$;  Prandtl number, $\mathrm{Pr}$, which is $\nu/\kappa$; 
\begin{eqnarray}
\textrm{Froude number }  \mathrm{Fr}  & = &   \frac{u_\mathrm{rms}/d}{N} = \frac{u_\mathrm{rms}}{d N}; \\\label{eq:Fr}
\textrm{Richardson  number }  \mathrm{Ri}  & = & \frac{N|b|_\mathrm{rms}d }{u_\mathrm{rms}^2} .\label{eq:Ri} 
\end{eqnarray}
Note that $\mathrm{Ri}  \approx  \mathrm{Fr} ^{-2}$.  For a detailed derivation, refer to Verma~\cite{Verma:book:BDF}.

Multiscale energy transfers are conveniently described in Fourier space. Here, the one-dimensional kinetic and potential energy spectra are defined as
\bea
E_u(k) =  \sum_{k-1 < k' \le k} \frac{1}{2} |{\bf u(k)}|^2;~~~E_b(k)  =  \sum_{k-1 < k' \le k} \frac{1}{2} |b{\bf(k)}|^2 .
\label{eq:SST_E_ub_k}
\eea
The nonlinear energy transfers across modes are quantified using energy fluxes or energy cascade rates.  The kinetic energy flux $\Pi_u(k_0)$ for a wavenumber sphere of radius $k_0$ is the total kinetic energy leaving the said sphere due to nonlinear interactions.  The potential energy flux $\Pi_{b}(k_0)$ is defined similarly.   Using the mode-to-mode kinetic/potential energy transfers, the corresponding fluxes are computed as
\bea
\Pi_{u}(k_0) & = & \sum_{|{\bf k'}|>k_0}   \sum_{|{\bf p}|>k_0}
 -\Im \left[  {\bf  \{  k' \cdot u(q) \} \{ u({\bf p}) \cdot u({\bf k'}) \} }  \right] ,
\label{eq:SST_KE_flux}  \\
\Pi_{b}(k_0) & = & \sum_{|{\bf k'}|>k_0}   \sum_{|{\bf p}|>k_0} 
-\Im \left[  {\bf  \{  k' \cdot u(q) \} } \{ b({\bf p})  b({\bf k'}) \}  \right],
\label{eq:SST_scalar_flux}
\eea
where ${\bf k'+p+q}=0$,  the giver Fourier modes (with wavenumbers ${\bf p}$) are within the sphere, while the receiver Fourier modes (with wavenumbers ${\bf k'}$) are outside the sphere.  See Dar et al.~\cite{Dar:PD2001} and Verma~\cite{Verma:PR2004,Verma:book:BDF} for details.

Under a steady state, in the inertial range where ${\bf F}_u = 0$ and the dissipative effects are negligible, the  equations for the kinetic and potential energies yield~\cite{Verma:NJP2017,Verma:book:BDF}
 \bea
  \frac{d}{dk} \Pi_u(k) & = & \mathcal{F}_B(k)  ,
  \label{eq:SST_energetics_steady_dPiu} \\
    \frac{d}{dk} \Pi_b(k) & = &- \mathcal{F}_B(k),
  \label{eq:SST_energetics_steady_dPirho}
  \eea
    where
\bea 
\mathcal{F}_B(k) & = & -\sum_{ |{\bf k'}| = k}  N  \Re [b({\bf k}') u_z^*({\bf k}')]
\label{eq:SST:F_B(k)}
\eea
is the energy supply rate to the kinetic energy due to buoyancy.  It is shown below that  $\mathcal{F}_B(k)  < 0$.   A sum of Eqs.~(\ref{eq:SST_energetics_steady_dPiu}, \ref{eq:SST_energetics_steady_dPirho}) yields 
\be
\Pi_u(k) + \Pi_b(k) = \mathrm{const},
\label{eq:SST_total_flux}
\ee   
or in the inertial range, the total energy flux is a constant.  For more details, refer to Verma~\cite{Verma:NJP2017, Verma:book:BDF}.

The nature of the SST crucially depends on the relative strength of the buoyancy and nonlinear term, or $\mathrm{Ri}$~\cite{Verma:book:BDF}.  For small $\mathrm{Ri}$, turbulence is similar to that of passive scalar turbulence, while for moderate $\mathrm{Ri}$,  buoyancy leads to $E_u(k) \sim k^{-11/5}$ (to be described in the subsequent subsections). The flow become strongly anisotropic and quasi-two-dimensional for strong buoyancy or large  $\mathrm{Ri}$~\cite{Davidson:book:TurbulenceRotating,Lindborg:JFM2006}.   

The phenomenology of SST with moderate stratification is described in the next two subsections,

\subsection{Energetics of moderately stratified turbulence}
\label{subsec:Energetics_SST}

For moderately stratified turbulence, the flow is nearly isotropic.   For such flows, an equation for the total kinetic  energy with ${\bf F}_u = 0$ and $\nu=0$ is
\bea
\dot{E}_u & = &-\la N b u_z \ra = \mathcal{F}_B
\eea
because $\la \nabla \cdot  [u^2 {\bf u}]  \ra = \la \nabla \cdot  [\sigma {\bf u}]  \ra=0$ for periodic or vanishing boundary condition.  Here $\la . \ra$ stands for an average over the real space.

An inviscid and nondiffusive stably stratified flow with ${\bf F}_u=0$ supports inertial gravity waves.  For such waves,  the  $\mathcal{F}_B =0$ with the kinetic and potential energies exchanging among themselves in a periodic manner.  This is a neutral state  for which the fluctuations neither grow nor decay.  Under an introduction of nonlinearity, $\mathcal{F}_B$ becomes negative.  If this were not the case, the kinetic energy of the system would grow and make the flow unstable.  Thus it is demonstrated that  $\mathcal{F}_B < 0$ for SST.  The above arguments are for the global $\mathcal{F}_B$, but it is reasonable to assume that for a generic stable system, $\mathcal{F}_B(k) < 0$.
 \begin{figure}[htbp]
\begin{center}
\includegraphics[scale = 0.6]{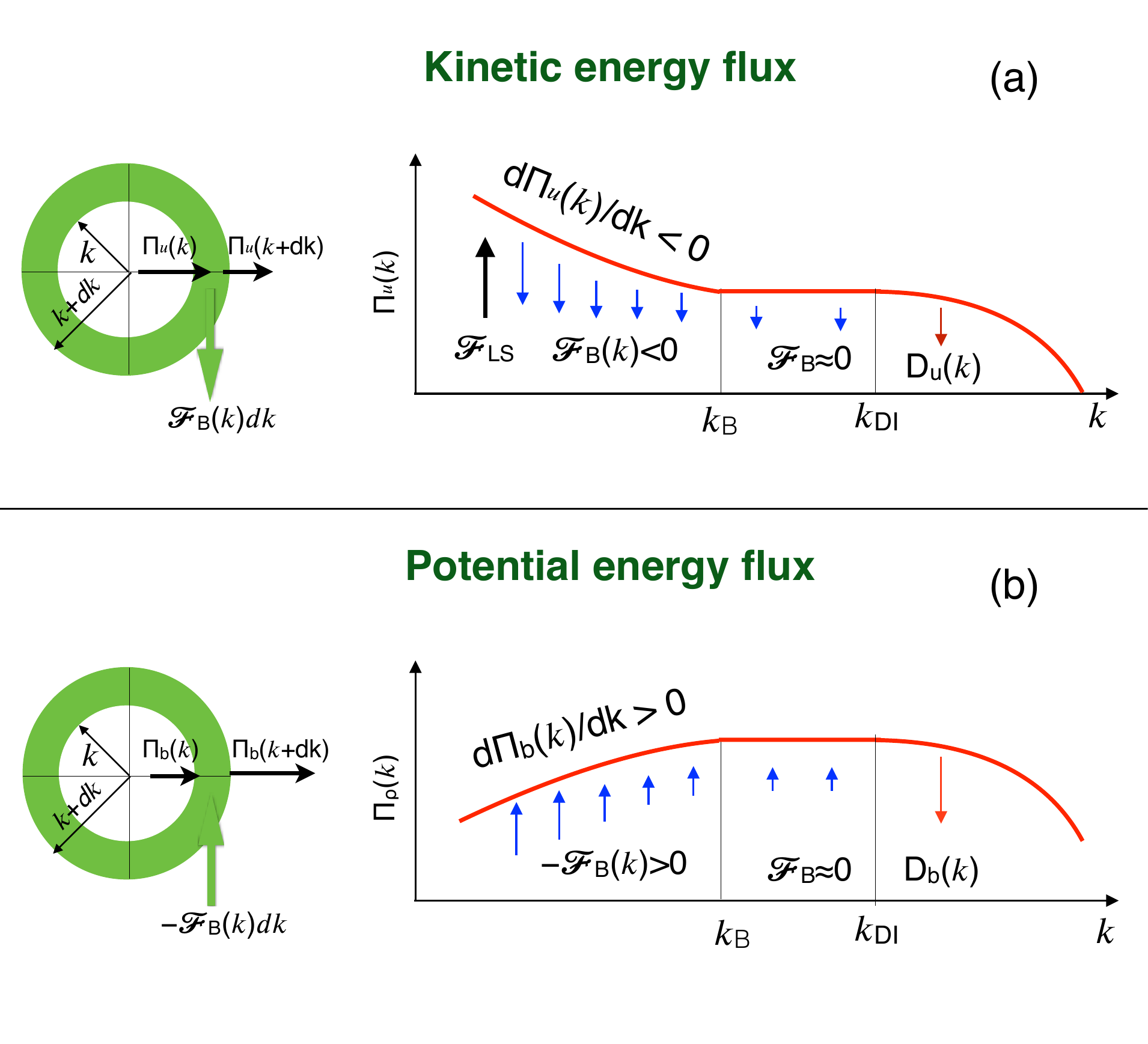}
\end{center}
\caption{ For SST: (a) A schematic diagram of the kinetic energy flux $\Pi_u(k)$.  In the band $k_f \ll k \ll k_B$, $d\Pi_u(k)/dk < 0$ because $\mathcal{F}_B(k) < 0$. However, for $k_B \ll k  \ll k_\mathrm{DI}$, $\mathcal{F}_B(k) \approx 0$, hence $\Pi_u(k) = \mathrm{const}$. The left subfigures illustrates the decreasing $\Pi_u(k)$ and negative $\mathcal{F}_B(k)$.  (b) Similar figures for the potential energy flux $\Pi_b(k)$ that increases in the band   $k_f \ll k \ll k_B$.  }
\label{fig:SST:Pi_schematic_SST}
\end{figure} 

For a negative $\mathcal{F}_B(k)$, Eq.~(\ref{eq:SST_energetics_steady_dPiu}) yields
 \be
 \frac{d}{dk} \Pi_u(k) < 0.
 \label{eq:SST_dPi_dk<0}
 \ee
 Thus, the kinetic energy flux $\Pi_u(k)$ of SST decreases with $k$. Consequently  $E_u(k)$ is steeper than the Kolmogorov's spectrum (for which $\Pi_u(k) \sim$ constant).   Using Eq.~(\ref{eq:SST_total_flux}), it is  deduced that the potential energy flux $\Pi_b(k)$ increases with $k$.  As a result, the scalar energy spectrum is shallower than $k^{-5/3}$.  These features are illustrated  in Fig.~\ref{fig:SST:Pi_schematic_SST}.

After this background,   Bologiano-Obukhov (BO) scaling \cite{Bolgiano:JGR1959, Obukhov:DANS1959} is presented in the next subsection. 

\subsection{Bolgiano-Obhukhov phenomenology for moderately stratified turbulence}
\label{subsec:BO_phenomenology}
  In BO phenomenology,  a force balance  between the nonlinear term and the buoyancy, as well as constancy of  potential energy flux yield
\bea
k u_k^2  & = & N b_k,   \label{eq:force_balance_b} \\
\Pi_b & = &  k b_k^2 u_k = \epsilon_b. \label{eq:SST_const_Pi_rho}
\eea
These equations yield the following spectra and fluxes:
 \begin{eqnarray}
E_u(k) & =  & c_1 \epsilon_b^{2/5}  N^{4/5} k^{-11/5}, \label{eq:Eu_2} \\
E_b(k) & =  & c_2 \epsilon_b^{4/5}  N^{-2/5} k^{-7/5}, \label{eq:Eb} \\
\Pi_u(k) & = & c_3  \epsilon_b^{3/5} N^{6/5} k^{-4/5},  \label{eq:pi_u} \\
\Pi_b(k) & = &  \epsilon_b. \label{eq:pi_b} 
\end{eqnarray}
 Clearly, $\Pi_u(k)$ decreases with $k$, consistent with Eq.~(\ref{eq:SST_dPi_dk<0}). 

Bolgiano \cite{Bolgiano:JGR1959} and Obukhov~\cite{Obukhov:DANS1959}  also argued that buoyancy weakens at large $k$ (before the start of dissipation range), hence yielding passive scalar turbulence like behaviour~\cite{Lesieur:book:Turbulence}:
\bea
E_u(k) & =  & K_{Ko}  \epsilon_u^{2/3}k^{-5/3}, \label{eq:Eu_KO2} \\
E_b(k) & =  & K_\mathrm{OC} \epsilon_u^{-1/3}\epsilon_b k^{-5/3}, \label{eq:Eb_KO} \\
\Pi_u(k) & = &  \epsilon_u,  \label{eq:pi_u_KO} \\
\Pi_b(k) & = &  \epsilon_b, \label{eq:pi_b_KO}
\end{eqnarray}
where $\epsilon_u$ is the viscous dissipation rate, and $K_\mathrm{Ko}, K_\mathrm{OC}$ are respectively Kolmogorov's and Obukhov-Corrsin's constants.  The behavioural transition from one regime to another  occurs near Bolgiano wavenumber $k_B$:
\be 
k_B \approx N^{3/2} \epsilon_u^{-5/4} \epsilon_b^{3/4}.
\label{eq:k_B}
\ee

In the following subsection it is shown that the latter regime (passive scalar turbulence) does not exist in SST with moderate stratification.    

\subsection{Revision of Bolgiano-Obhukhov phenomenology }
Let us start with  the conservation law, Eq.~(\ref{eq:SST_total_flux}):
\be
(k u_k^3 + k b_k^2 u_k  = \epsilon) \Rightarrow k u_k^3 \left[ 1 +  \frac{k^2 u_k^2}{N^2} \right]= \epsilon.
\label{eq:flux_sum2}
\ee
A numerical solution of the above equation shows that $\Pi_b \gg \Pi_u$~\cite{Alam:arxiv2018}.  More importantly, the solution does not exhibit any transition from BO scaling [Eqs.~(\ref{eq:Eu_2}-\ref{eq:pi_b})] to passive-scalar scaling [Eqs.~(\ref{eq:Eu_KO2}-\ref{eq:pi_b_KO})]. This is essentially because at small $k$, $u_k$ is too small  to initiate a constant kinetic energy flux.   

A  proof for the absence of the above transition is as follows. In the second regime ($k > k_B$), buoyancy should be much smaller than the the nonlinear term.   Let us estimate the ratio of the buoyancy and the nonlinear term in the second regime: 
\bea
\frac{N b_k}{k u_k^2} & \approx &
\frac{N \epsilon_b^{1/2} \epsilon_u^{-1/6} k^{-1/3}}{k \epsilon_u^{2/3} k^{-2/3}}  \approx  N \epsilon_b^{1/2} \epsilon_u^{-5/6} k^{1/3}.
\eea
 Since $k >k_B$, using Eq.~(\ref{eq:k_B}), we deduce that
 \bea
\frac{N b_k}{k u_k^2} > N \epsilon_b^{1/2} \epsilon_u^{-5/6} k_B^{1/3} \approx  N^{3/2} \epsilon_b^{3/4} \epsilon_u^{-5/4}
\gg 1
\eea
because $\epsilon_b \gg \epsilon_u$.  Therefore, buoyancy should dominate in the second regime.  This is a contradiction.  Therefore, we prove using contradiction that the second regime does not exist.  We call the above phenomenology as revised BO phenomenology.

In the next subsection we will describe numerical simulations that verify  BO phenomenology.

\subsection{Numerical results on SST}
\label{sec:SST_num} 

There are a large number of simulations on SST with strong stratification~(\cite{Lindborg:JFM2006,Davidson:book:TurbulenceRotating,Vallgren:PRL2011} and references therein), and only a handful  on SST with moderate stratification ($\mathrm{Ri} \approx 1$)~\cite{Kimura:JFM1996, Kumar:PRE2014, Rosenberg:PF2015}.  Here we present the results of Kumar {\em et al.}~\cite{Kumar:PRE2014} because it shows conclusive evidences in favour of revised BO scaling.

Kumar {\em et al.}~\cite{Kumar:PRE2014}  performed a numerical simulation of SST for $\mathrm{Pr}=1$ on a $1024^3$ grid.  They employed forcing at large scales to attain a steady state.  The steady flow has Reynolds number $\mathrm{Re}\approx 649$, and Richardson number  $\mathrm{Ri} \approx0.01$.  The flow is nearly isotropic because $\mathrm{Ri}$ is not too far from unity.   
Kumar {\em et al.} reported that the  $E_u(k) \sim k^{-11/5}$ provides a  fit to the numerical data that $k^{-5/3}$ spectrum.  Similarly, the potential energy spectrum is better described by $k^{-7/5}$ spectrum than $k^{-5/3}$.    They also reported that $\mathcal{F}_B(k) < 0$, $\Pi_\theta \approx \mathrm{const}$, while $\Pi_u(k)$  decreases with $k$.   Note that they do not observe any crossover from $k^{-11/5}$ to $k^{-5/3}$. This feature may be due to relatively lower resolution of the numerical simulation, or due to absence of crossover as predicted by revised BO scaling~\cite{Alam:arxiv2018}.    Thus, the numerical results of  Kumar {\em et al.}~\cite{Kumar:PRE2014}  are in general  agreement with   BO phenomenology or revised BO phenomenology.    The fluxes of kinetic and potential energies need to be recomputed using higher-resolution simulations for a more accurate verification.

The  phenomenological arguments for turbulent thermal convection is presented in the next section, . 

\section{Turbulent thermal convection}
\label{sec:turbulent convection} 

In thermal convection, density is unstably stratified as shown in Fig.~\ref{fig:density_schematic}(b).  This section focuses on a special class of thermal convection called {\em Rayleigh-B\'{e}nard convection} (RBC).

\subsection{Formalism}
In RBC, a fluid is confined between two conducting horizontal plates that are kept at $z=0$ and $d$; the temperatures of these plates are $T_b$ and $T_t$ respectively with $T_b > T_c$.     The local temperature is a superposition of externally-imposed linearly varying   temperature  $\bar{T}(z)$ and fluctuation $\theta(x,y,z)$:
\be
T(x,y,z) = \bar{T}(z) + \theta(x,y,z),
\label{eq:T}
\ee
where
\be
\bar{T}(z) = T_b + \frac{d\bar{T}}{dz} z = T_b - \frac{T_b-T_t}{d} z.
\ee

For the above system, the equations of motion for the velocity and temperature fluctuations under Boussinesq approximation are
 \bea
 \frac{\partial {\bf u}}{\partial t} + ({\bf u} \cdot \nabla) {\bf u} & = &  -\frac{1}{\rho_m} \nabla \sigma   + \alpha g \theta \hat{z} + \nu \nabla^2 {\bf u}, \label{eq:RBC1}
 \\
\frac{\partial \theta}{\partial t} +  ({\bf u} \cdot \nabla)  \theta & = &  \frac{\Delta}{d} u_z +\kappa \nabla^2 \theta, \label{eq:RBC2} \\
\nabla \cdot {\bf u} & = & 0, \label{eq:RBC3}
\eea 
\label{eq:basic_eqns}
where $\Delta = T_b-T_t$, and $\nu, \kappa$ are the kinematic viscosity and {\em thermal diffusivity} respectively~\cite{Verma:book:BDF}.   Two nondimensional parameters of RBC are Rayleigh number $\mathrm{Ra}$ and Prandtl number $\mathrm{Pr}$ that are defined as
\be
\mathrm{Ra}  =  \frac{\alpha g d^3 \Delta }{\nu \kappa};~~\mathrm{Pr} =  \frac{\nu}{\kappa}.
\ee

Under the inviscid and nondiffusive limit,  it can be shown that 
\be
\frac{1}{2} \int d{\bf r} \left( u^2   -   \frac{\alpha gd}{\Delta}  \theta^2 \right)
\label{eq:turbulent convection_conservation_real}
\ee
is conserved for RBC.  The scalar quantity for thermal convection is $\theta$ (corresponding to $b$ of SST);  one-dimensional   spectrum for the temperature fluctuation ($\theta$) is defined as
\bea
E_\theta(k)  =  \sum_{k-1 < k' \le k} \frac{1}{2} |\theta{\bf(k)}|^2 ,
\label{eq:RBC_E_ub_k}
\eea
and the temperature flux (for nonlinear scale-by-scale transfer of $\theta^2$) is
\bea
\Pi_{\theta}(k_0) & = & \sum_{|{\bf k'}|>k_0}   \sum_{|{\bf p}|>k_0} 
-\Im \left[  {\bf  \{  k' \cdot u(q) \} } \{ \theta({\bf p}) \theta({\bf k'}) \}  \right].
\label{eq:RBC_scalar_flux}
\eea
The definitions of kinetic energy spectrum and flux are same as those in Sec.~\ref{sec:SST}.

Under a steady state, in the inertial range where the dissipation and diffusion effects are negligible, the  energy equations yield~\cite{Verma:NJP2017, Verma:book:BDF}
 \bea
\frac{d}{dk} \Pi_u(k) = \mathcal{F}_B(k) ,   
\label{eq:RBC_Pi_u_steady_state} \\
\frac{d }{dk} \Pi_\theta(k) = \frac{\Delta}{\alpha g d}  \mathcal{F}_B(k),  
\label{eq:RBC_Pi_theta_steady_state}
\eea
where 
\be
\mathcal{F}_B(k) = \sum_{|{\bf k'}|>k_0} \alpha g  \Re [\theta({\bf k}) u_z^*({\bf k})]
\ee
is the kinetic energy feed by buoyancy.  Using Eqs.~(\ref{eq:RBC_Pi_u_steady_state}, \ref{eq:RBC_Pi_theta_steady_state}) it is  deduced that
\be
\Pi_u(k)  -  \frac{\alpha gd}{\Delta} \Pi_\theta(k) = \mathrm{const} = C_1.
\label{eq:RBC:RBC_flux_diff}
\ee
The above equation can be transformed to the following:
\be
\frac{U^3}{d} \left[  \Pi'_u(k)-   \frac{\alpha g\Delta d}{U^2} \Pi'_\theta(k) 
\right]  \approx C_1,
\ee
where $ \Pi'_u(k) = \Pi_u(k)/(U^3/d)$ and $\Pi'_\theta(k) = \Pi_\theta(k)/(U\Delta^2/d)$ are  the nondimensional kinetic energy flux and temperature flux respectively.   Since $U \approx \sqrt{\alpha g\Delta d}$, we deduce that
\be
\Pi'_u(k) -  \Pi'_\theta(k) \approx \mathrm{const} = C_2.
\label{eq:RBC:RBC_flux_diff2}
\ee
Verma~\cite{Verma:book:BDF} showed that $C_2 \approx 0$.  
 
%


The boundary layers near the thermal plates affect the temperature field.  Pandey and Verma~\cite{Pandey:PF2016}, and Verma {\em et al.}~\cite{Verma:NJP2017} showed that the mean temperature profile is $\theta_m(z)  \approx z-1/2$, whose Fourier transform and temperature spectrum are
\be
{\theta}_m(0,0,k_z)  \approx  -\frac{1}{\pi k_z}  \Rightarrow E_{\theta_m}(k)  \sim k^{-2}
\ee
 for even $k_z$, and zero otherwise.

After this background, I describe a phenomenology of  turbulent thermal convection that yields the spectra and fluxes of the velocity and temperature fields.

\subsection{Phenomenology of turbulent thermal convection}
\label{subsec:TTC_phenomenology}
As described in the introduction, motivated by  similarities between the equations of stably stratified flows and  thermal convection, many researchers argued that  BO phenomenology would also apply to  turbulent thermal convection~(\cite{Ahlers:RMP2009,Chilla:EPJE2012,Lohse:ARFM2010,Lvov:PRL1991,Lvov:PD1992,Rubinstein:NASA1994,Verma:book:BDF} and references therein).  It is shown below that the nature of energy feed by buoyancy in SST and  turbulent convection are very different.   Based on these new energetics arguments and several properties of turbulent convection, it is shown that $E_u(k) \sim k^{-5/3}$ and $\Pi_u \approx $ const.

 \begin{figure}[htbp]
\begin{center}
\includegraphics[scale = 0.5]{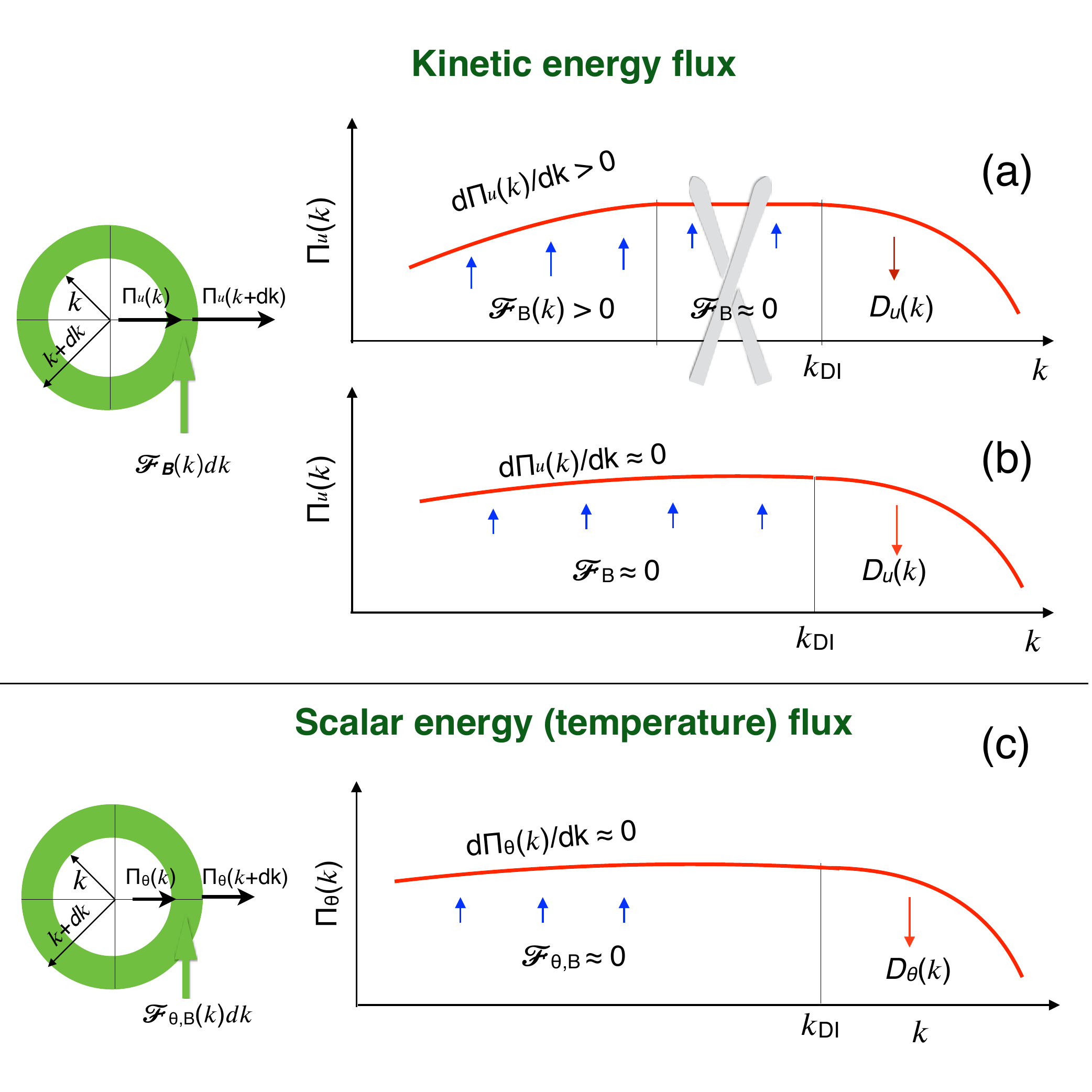}
\end{center}
\caption{For turbulent thermal convection: (a) A schematic diagram of  $\Pi_u(k)$ for which $\mathcal{F}_B(k) > 0$. (b) Equation~(\ref{eq:RBC:fluxes}) however reveals that $\Pi_u(k) \approx \mathrm{const}$.  (c) A schematic diagram of  $\Pi_\theta(k)$ with positive feed from buoyancy, but Eq.~(\ref{eq:RBC:fluxes}) shows that $\Pi_\theta(k) \approx \mathrm{const}$. }
\label{fig:Pi_schematic_RBC}
\end{figure} 

 Since hot plumes ascend and cold plumes descend,   $\theta$ and $u_z$ are positively correlated, or 
\be
\mathcal{F}_B  = \langle \theta({\bf r}) u_z({\bf r}) \rangle > 0.
\ee
Hence it is expected that  
\be
\mathcal{F}_B(k) = \sum_{k-1 < k' \le k} \alpha g \langle \theta({\bf k'}) u_z^*({\bf k'}) \rangle > 0.
\ee
Therefore, using Eqs.~(\ref{eq:RBC_Pi_u_steady_state}, \ref{eq:RBC_Pi_theta_steady_state}) it is deduced that 
\be
 \frac{d}{dk} \Pi_u(k) >0;~~~~ \frac{d}{dk} \Pi_\theta(k) >0.
 \ee
Hence, both $\Pi_u(k)$ and $\Pi_\theta(k)$ are expected to increase with $k$.  A recent numerical simulation of Verma {\em et al.}~\cite{Verma:NJP2017} however reveal that $\mathcal{F}_B(k)  \sim k^{-5/3}$.  Hence
\be
\mathcal{F}_B(k) = \alpha g \langle \theta({\bf k}) u_z^*({\bf k}) \rangle   \sim (\alpha g U \Delta ) k^{-5/3},
\ee
where  $U \sim \sqrt{\alpha g U \Delta}$, and $k$ is a nondimensionalized variable (normalized with $1/d$). In the inertial range where $k \gg 1$, we expect   $\mathcal{F}_B(k) \sim k^{-5/3} \rightarrow 0$.     Therefore, it  can be claimed that in the inertial range,
  \be
  \Pi_u(k)   \approx \mathrm{const};~~~\Pi_\theta(k)  \approx \mathrm{const}.
  \label{eq:RBC:fluxes}
  \ee 
 
Constancy of kinetic energy flux prompts us to  predict that $E_u(k)$ of turbulent convection follows Kolmogorov's spectrum:
\be
E_u(k) = K_\mathrm{Ko} (\Pi_u)^{2/3} k^{-5/3}.
\ee
These observations are consistent with (a)  Verma's~\cite{Verma:book:BDF} finding that  $\mathrm{Ri} \approx 0.1$ for turbulent convection; (b) Pandey and Vema's~\cite{Pandey:PF2016} results that turbulent convection is driven primarily by the pressure gradient, as in 3D hydrodynamics.   Thus, buoyancy essentially  supplies kinetic energy at  large scales, and it does not affect the inertial range spectrum significantly.   Recent numerical simulations of Kumar {\em et al.}~\cite{Kumar:PRE2014}, Verma {\em et al.}~\cite{Verma:NJP2017}, and Verma~\cite{Verma:book:BDF} verify the above conjecture (see Sec.~\ref{sec:turbulent convection_num}). Note that  these arguments rule out BO scaling for turbulent convection.

The temperature spectrum $E_\theta(k)$ however is not proportional to $k^{-5/3}$ due to the walls (though $\Pi_\theta(k)  \approx \mathrm{const}$).   As described in the previous subsection, the mean temperature $\theta_m(z)$ exhibits $k^{-2}$ spectrum.  In addition, the fluctuating part of $\theta$ generate another  branch in $E_\theta(k)$.  These two branches however  yield a constant $\Pi_\theta(k)$.

The numerical results on turbulent convection are presented in the next subsection.

\subsection{Numerical verification  of Kolmogorov-like scaling in turbulent convection}
\label{sec:turbulent convection_num} 

Many researchers performed numerical simulations and experiments of turbulent thermal convection to test whether it follows Kolmogorov-like scaling or BO scaling.  The results were somewhat inconclusive till recently due to various reasons~(\cite{Ahlers:RMP2009,Chilla:EPJE2012,Lohse:ARFM2010,Lvov:PRL1991,Lvov:PD1992,Niemela:Nature2000,
Rubinstein:NASA1994,Skrbek:PRE2002,Verma:book:BDF} and references therein).   Most experiments measure the temperature and/or velocity fields at set of physical locations.  Hence, determination of $E_u(k)$  requires invocation of Taylor's hypothesis to convert the frequency spectrum $E_u(f)$ to $E_u(k)$.  However, lack of a mean flow may invalidate such transformation \cite{Kumar:RSOS2018}, which may be the reason for variations in the experimental results.

Here, due to lack of space, only some of the numerical results are listed.  Grossmann and Lohse~\cite{Grossmann:PRL1991} simulated RBC using  small number of truncated Fourier-Weierstrass modes and obtain Kolmogorov's scaling.  Borue and Orszag~\cite{Borue:JSC1997}, Skandera {\em et al.}~\cite{Skandera:CP2007}, and Kerr~\cite{Kerr:JFM1996} reported $k^{-5/3}$ spectrum for the velocity and temperature field.  Note however that  Borue and Orszag~\cite{Borue:JSC1997} and Skandera {\em et al.}~\cite{Skandera:CP2007} employed periodic boundary conditions for their simulations.  Though some of the above simulations report that $E_u(k) \sim k^{-5/3}$, these results were not fully convincing.  Recently Kumar {\em et al.}~\cite{Kumar:PRE2014} and Verma {\em et al.}~\cite{Verma:NJP2017} performed numerical simulation of turbulent thermal convection and concluded that BO scaling is ruled out for turbulent convection, and that turbulent convection has behaviour similar to hydrodynamic turbulence. 

Verma et al.~\cite{Verma:NJP2017} simulated turbulent thermal convection  for $\mathrm{Pr} = 1$ and $\mathrm{Ra} = 1.1 \times 10^{11}$ on a $4096^3$ grid.  They observed that $E_u(k) \sim k^{-5/3}$ with a constant flux, consistent with the phenomenology described in Sec.~\ref{subsec:TTC_phenomenology}.  They also reported that the temperature spectrum $E_\theta(k)$   exhibits bi-spectrum---the upper branch varies as $k^{-2}$, while lower branch deviates from both $k^{-5/3}$ and $k^{-7/5}$. Thus, the temperature spectrum is far from predictions of passive scalar turbulence or BO phenomenology.  Yet, the temperature spectrum $\Pi_\theta(k) \sim \mathrm{const}$, and $\Pi_u(k) - \Pi_\theta(k)$, consistent with the conservation law of Eq.~(\ref{eq:RBC:RBC_flux_diff2}).  Also, $\mathcal{F}_B(k) > 0$ consistent with the phenomenology of Sec.~\ref{subsec:TTC_phenomenology}.


In addition, Verma {\em et al.}~\cite{Verma:NJP2017} showed that the shell-to-shell energy transfers are local and forward, similar to those in three-dimensional hydrodynamic turbulence. Also, Nath {\em et al.}~\cite{Nath:PRF2016} showed that the flow of a turbulent convection is nearly isotropic.  The shell model of turbulent convection too shows similar results as above~\cite{Kumar:PRE2015}. Thus,  the turbulence properties of  turbulent thermal convection are similar to hydrodynamic turbulence.   This is a very useful result, and it enables us to employ the turbulence models of hydrodynamic turbulence to turbulent thermal convection.   Vashistha {\em et al.}~\cite{Vashishtha:PRE2018} exploited this observation and performed large-eddy simulation (LES)  of turbulent convection using the hydrodynamic LES.

Next section contains a summary of the main differences between SST and thermal convection.

\section{Main differences between SST and turbulent thermal convection: A summary}
\label{sec:diff} 

In this paper the properties of SST (with focus on moderate stratification) and turbulent convection are contrasted, and it is shown that some of the important turbulence properties of these systems are very different due to the nature of stratification.  The former  is stable, while the latter  is unstable. This feature leads to differences  in the signs of  the kinetic energy feed by buoyancy, $\mathcal{F}_B(k)$.  Such energy feeds have profound influence on the energy spectrum and flux. 

The differences between SST and turbulent thermal convection are summarised in Table~\ref{tab:contrast}.  SST are classified into three classes: $\mathrm{Ri} \ll 1$, $\mathrm{Ri} \sim 1$, and $\mathrm{Ri} \gg 1$.  In this paper the focus  is on SST with $\mathrm{Ri} \sim 1$, yet for completeness,   the properties of $\mathrm{Ri} \ll 1$ and $\mathrm{Ri} \gg 1$ are also summarised in the table.  The flows with $\mathrm{Ri} \ll 1$  are described by passive scalar turbulence in which the velocity field follows Kolmogorov's $k^{-5/3}$ spectrum, and  the density field is advected by the velocity field as a passive scalar.  It has been shown that the passive scalar too yields $k^{-5/3}$ spectrum~\cite{Lesieur:book:Turbulence,Verma:IJMPB2001}.  These spectra arise because the buoyancy is very weak for this case.

The flows with $\mathrm{Ri} \gg 1$  (SST with strong stratification) is not  well understood, in particular the columns with entries ``?" in the Table.  For  such flows, there are some works to compute the spectra and fluxes, as well as $\mathcal{F}_B(k)$~\cite{Lindborg:JFM2006,Vallgren:PRL2011,Davidson:book:TurbulenceRotating}.  Yet, it is safe to say that there is no convergence on  topics such as fluxes.   For example, Vallgren {\em et al.}~\cite{Vallgren:PRL2011} observed that the total energy flux is positive, but we are not aware of rigorous computation of $\Pi_u(k)$ and $\Pi_b(k)$ individually.  These fluxes would have important contributions on the construction of phenomenological theories of such systems.  Similarly, for this is case, $\mathcal{F}_B(k)$  could be positive (contrast to other two cases) due to two dimensionality.   For more details on this regime, refer to~\cite{Lindborg:JFM2006,Vallgren:PRL2011,Davidson:book:TurbulenceRotating,Verma:book:BDF}
 
The energetics arguments described in the paper are quite general, and they can be employed to other related flows.  For example, Rayleigh-Taylor turbulence, which is unstably stratified, too shows properties similar to hydrodynamic turbulence~\cite{Boffetta:ARFM2016,Verma:book:BDF}.  For further details on  Rayleigh-Taylor instability, refer to Abarzhi~\cite{Abarzhi:EPL2010, Abarzhi:PTRS2010}.

The aforementioned arguments are quite robust and promising, and they are in  good agreement with simulation results.   Yet, there are many unresolved issues.  For example, the physics of  SST with strong stratifictio ($\mathrm{Ri} \gg 1$) is not well understood.  In addition, generalisation of the energetics arguments to two-dimensional flows, to boundary layers in thermal convection, and to extreme Prandtl numbers are yet to be satisfactorily performed.  

\addcontentsline{toc}{section}{Acknowledgments}
\ack
This paper is an expanded version (with some new topics) of the talk given at  the conference {\em Turbulence Mixing and Beyond 2017} organized at ICTP Trieste.  I thank Snezhana Abarzhi and ICTP for hosting this interesting meeting.   The numerical simulations presented in the paper were performed by Abhishek Kumar, to whom I am grateful.  I am also thankful to K. R. Sreenivasan, J\"{o}rg Schumacher, Jayant Bhattacharjee, Joe Niemela, L. Skrbek, Abhishek Kumar, Ambrish Pandey, Anirban Guha, Shadab Alam,  Shashwat Bhattacharya, and other members of our turbulence group for useful discussions and idea exchanges.   The  simulations were performed on Shaheen II of the Supercomputing Laboratory at King Abdullah University of Science and Technology (KAUST) under the project K1052,  on Chaos supercomputer of Simulation and Modeling Laboratory (SML), IIT Kanpur, and on HPC2013 of IIT Kanpur.  This work was supported by the research grant PLANEX/PHY/2015239 from Indian Space Research Organisation, India.

\pagestyle{empty}
\begin{landscape}
 \begin{table}[h]
 \caption{Contrasting  stably stratified turbulence (SST)  and turbulent thermal convection. SST is classified into three classes: $\mathrm{Ri} \ll 1$, $\mathrm{Ri} \sim 1$, and $\mathrm{Ri} \gg 1$.  In the table, "?" means that the particular issue remains largely unresolved.  Primary references are cited in the first row, but for more details refer to Sec.~\ref{sec:diff}. }
  \centering
  {\begin{tabular}{ c || c| c|c ||c } \hline\hline
\multirow{2}{*}{Property} & \multicolumn{3}{c||}{SST}  &  Thermal convection  \\  \cline{2-5} 
 		& $\mathrm{Ri} \ll 1$~\cite{Kumar:PRE2014} & $\mathrm{Ri} \sim 1$~\cite{Bolgiano:JGR1959,Obukhov:DANS1959, Kumar:PRE2014}  & $\mathrm{Ri} \gg 1$~\cite{Lindborg:JFM2006,Davidson:book:TurbulenceRotating} & $\mathrm{Ri} \approx 0.1$~\cite{Kumar:PRE2014,Verma:NJP2017}  \\ \hline 
Froude no & $\gg 1$ & $\sim 1$ & $\ll 1$ & - \\ \hline		
Stability & \multicolumn{3}{c||}{Stable} & Unstable \\ \hline
Linear mode & \multicolumn{3}{c||}{Internal gavity waves} & Convective rolls \\ \hline
Isotropy & Nearly isotropic  &Nearly isotropic & Anisotropic & Nearly isotropic  \\ \hline
Phenomenology & passive scalar   &Bolgiano-Obukhov & quasi-two-dimensional & Kolmogorov-like   \\ 
	& turbulence & && \\ \hline
$\mathcal{F}_B(k)$ & Negative &Negative & ?&Positive  \\ \hline
$\Pi_u(k)$ &  $\Pi_u(k) \approx$ const &  $\Pi_u(k)$ decreases with $k$ & ? &  $\Pi_u(k) \approx $ const \\ \hline
\multirow{2}{*}{$E_u(k)$} & \multirow{2}{*}{$k^{-5/3}$}  & \multirow{2}{*}{$k^{-11/5}$} & $E_\perp(k_\perp) \sim k_\perp^{-5/3}$ & \multirow{2}{*}{$k^{-5/3}$}  \\  
 & 	& & $E(k_\parallel) \sim k_\parallel^{-5/3}$	&     \\  \hline
Scalar flux &  $\Pi_b(k) \sim \mathrm{const}$ &   $\Pi_b(k) \sim \mathrm{const}$ &  ? 
& $\Pi_\theta(k) \sim  \mathrm{const}$  \\ \hline
 Scalar & \multirow{2}{*}{$E_b(k) \sim k^{-5/3}$} &  \multirow{2}{*}{$E_b(k) \sim k^{-7/5}$} & ? & Bi-spectrum with  $k^{-2}$ and \\ 
spectrum &&&& another branch for fluctuations.   \\ \hline
Conserved flux & \multicolumn{3}{c||}{$\Pi_u(k) + \Pi_b(k)$} & $\Pi_u(k) - \Pi_\theta(k)$  \\ \hline
Examples & Atmosphere of  & Parts of oceans & Earth's atmosphere & Earth's boundary layer \\
 & small planets &  &  &  \\ \hline
\end{tabular}}
\label{tab:contrast}
\end{table}
\end{landscape}
\pagestyle{plain}

 {\bf References} \\


\end{document}